\documentclass[10pt,conference]{IEEEtran}\IEEEoverridecommandlockouts
\usepackage{epsfig,graphicx,subfigure,psfrag,amsmath,cases,bm}
\usepackage{latexsym,amssymb,amsmath,epsfig,subfigure,algorithm,mathtools}
\usepackage{algorithmic}
\usepackage{color}
\usepackage{url}
\usepackage{scrtime}
\usepackage[left=0.6in,top=0.40in,bottom=0.52in,right=0.6in]{geometry}
\author{\IEEEauthorblockN {Yan Sun\IEEEauthorrefmark {1}, Derrick Wing Kwan Ng\IEEEauthorrefmark {2}, and Robert Schober\IEEEauthorrefmark {1}}

\IEEEauthorrefmark {1}Institute for Digital Communications, Friedrich-Alexander-University
Erlangen-N\"urnberg (FAU), Germany\\
\IEEEauthorrefmark {2}School of Electrical Engineering and Telecommunications, The University
of New South Wales, Australia\vspace*{-6mm}

\thanks{This work was supported in part by the AvH Professorship Program of the Alexander von Humboldt Foundation. Derrick Wing Kwan Ng is supported under Australian Research Council's Discovery Early Career Researcher Award funding scheme (project
number DE170100137).
}
}

\title{Resource Allocation for Secure  Full-Duplex \\ OFDMA Radio Systems\vspace*{-2mm}}

\newtheorem{Thm}{Theorem}

\newtheorem{T-Prob}{Transformed Problem}
\newtheorem{Prop}{Proposition}

\DeclareMathOperator{\Tr}{Tr}
\DeclareMathOperator{\Rank}{Rank}
\DeclareMathOperator{\maxo}{maximize}
\DeclareMathOperator{\mino}{minimize}

 \newcommand{\qed}{\hfill \ensuremath{\blacksquare}}

\newcommand{\abs}[1]{\lvert#1\rvert}
\newcommand{\norm}[1]{\lVert#1\rVert}
\begin{document}
\IEEEspecialpapernotice{(Invited Paper)}
\maketitle

\begin{abstract}
In this paper, we study the resource allocation for an orthogonal frequency division multiple access (OFDMA) radio system employing a full-duplex base station for serving multiple half-duplex downlink and uplink users simultaneously.
The resource allocation design objective is the maximization of the weighted system throughput while limiting the information leakage to guarantee secure simultaneous downlink and uplink transmission in the presence of potential eavesdroppers.
The algorithm design leads to a mixed combinatorial non-convex optimization problem and obtaining the globally optimal solution entails a prohibitively high computational complexity. Therefore, an efficient successive convex approximation based suboptimal iterative algorithm is proposed. Our simulation results confirm that the proposed suboptimal algorithm achieves a significant performance gain compared to two baseline schemes.
\end{abstract}
\renewcommand{\baselinestretch}{0.94}
\large\normalsize
\vspace*{-1mm}
\section{Introduction}
Secrecy and privacy are  critical concerns for the design of wireless communication systems due to the broadcast nature of the wireless medium \cite{Chen2015Multi}. Physical layer security is a new approach for preventing eavesdropping in future wireless communication systems \cite{ng2011secure}\nocite{li2013spatially}--\cite{Kwan14Robust}. Particularly, the base station (BS) can transmit artificial noise (AN) in the downlink (DL) to impair the information reception at potential eavesdroppers.
In \cite{ng2011secure}, a power allocation algorithm for maximizing the secrecy outage capacity via AN generation in orthogonal frequency division multiple access (OFDMA) relay systems was proposed.
In \cite{li2013spatially}, joint transmit signal and AN covariance matrix optimization was studied for secrecy rate maximization.
The authors of \cite{Kwan14Robust} developed a robust resource allocation algorithm  to guarantee DL communication security in multiuser communication systems.
However, the above works focus on ensuring secure DL transmission in half-duplex (HD) systems. The resulting schemes are not able to secure uplink (UL) transmission.

On the other hand, full-duplex (FD) transceivers allow simultaneous DL and UL transmission in the same frequency band \cite{sun2016optimalJournal}. Motivated by this property of FD, in \cite{zhu2014joint}\nocite{Sun16FDSecurity}--\cite{sun17WSAsecureFD}, an FD BS simultaneously protects DL and UL communication by transmitting AN in the DL to interfere potential eavesdroppers. We note that securing the UL is not possible with a conventional HD BS.
In \cite{zhu2014joint}, the joint design of information beamforming and AN generation for an FD BS was investigated to guarantee DL and UL communication security.
In \cite{Sun16FDSecurity}, the authors studied the tradeoff between the total DL transmit power consumption and the total UL transmit power consumption in secure multiuser FD systems.
The authors of \cite{sun17WSAsecureFD} proposed a suboptimal resource allocation algorithm for the maximization of the system secrecy throughput in FD systems.
However,  only single-carrier systems were considered in \cite{zhu2014joint}\nocite{Sun16FDSecurity}--\cite{sun17WSAsecureFD}, whereas today's wireless networks employ multicarrier transmission, e.g. the 4-th generation wireless communication systems (long-term evolution (LTE)) are based on OFDMA. Unfortunately, the resource allocation schemes proposed in \cite{zhu2014joint}\nocite{Sun16FDSecurity}--\cite{sun17WSAsecureFD} cannot be directly applied to FD OFDMA systems. In particular, the pairing of the DL and UL users on each subcarrier is a vital problem for FD OFDMA systems but was not considered in \cite{zhu2014joint}\nocite{Sun16FDSecurity}--\cite{sun17WSAsecureFD}. In fact, to the best of our knowledge, the resource allocation for secure FD OFDMA systems has not been investigated yet.

In this paper, we address the above issues. To this end, the resource allocation algorithm design for FD OFDMA systems is formulated as a non-convex optimization problem for the maximization of the weighted system throughput.  The maximum tolerable data rates for information leakage to potential eavesdroppers are limited for guaranteeing secure DL and UL transmission. Unfortunately, this optimization problem is in general intractable and obtaining the globally optimal solution may result in an unacceptably high computational complexity. Therefore, we develop a suboptimal resource allocation algorithm based on successive convex approximation to strike a balance between computational complexity and optimality.

\vspace*{-1mm}
\section{System Model}
In this section, we present the considered FD OFDMA wireless communication system model.

\vspace*{-1mm}
\subsection{Notation}%
We use boldface capital and lower case letters to denote matrices and vectors, respectively.   $\Tr(\mathbf{A})$ denotes the trace of matrix $\mathbf{A}$; $\mathbf{A}\succeq \mathbf{0}$ and $\mathbf{A}\preceq \mathbf{0}$ indicates that $\mathbf{A}$ is a positive semidefinite matrix and a negative semidefinite matrix, respectively; $\mathbf{A}^{-1}$ represents the inverse of matrix $\mathbf{A}$; $\mathbf{I}_N$ is the $N\times N$ identity matrix; $\mathbb{C}$ denotes the set of complex values; $\mathbb{C}^{N\times M}$ denotes the set of all $N\times M$ matrices with complex entries; $\mathbb{C}^{N\times 1}$ and $\mathbb{R}^{N\times 1}$ denote the sets of all $N\times 1$ vectors with complex and real entries, respectively; $\mathbb{H}^N$ denotes the set of all $N\times N$ Hermitian matrices; $\abs{\cdot}$ and $\norm{\cdot}$ denote the absolute value of a complex scalar and the Euclidean vector norm, respectively; ${\cal E}\{\cdot\}$ denotes statistical expectation;  $[x]^+$ stands for $\mathrm{max}\{0,x\}$; the circularly symmetric complex Gaussian distribution with mean $\mu$ and variance $\sigma^2$ is denoted by ${\cal CN}(\mu,\sigma^2)$; and $\sim$ stands for ``distributed as"; $\nabla_{\mathbf{x}} f(\mathbf{x})$ denotes the gradient vector of function $f(\mathbf{x})$ whose components are the partial derivatives of $f(\mathbf{x})$.

\vspace*{-1mm}
\subsection{FD OFDMA System Model}%
\begin{figure}
\centering\vspace*{-0mm}
\includegraphics[width=3in]{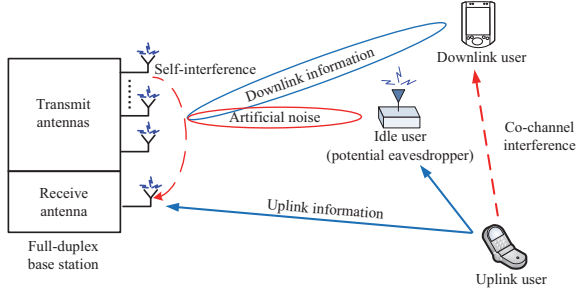}\vspace*{-3mm}
\caption{An OFDMA system with an FD BS, $K=1$ HD DL user, $J=1$ HD UL user, and $M=1$ HD idle user (potential eavesdropper).}
\label{fig:system_model}\vspace*{-5mm}
\end{figure}

We consider an FD OFDMA system which consists of an FD BS, $K$  DL users, $J$  UL users, and $M$ idle users, cf. Figure \ref{fig:system_model}.
The entire frequency band of $W$ Hertz is partitioned into ${N_{\mathrm{F}}}$ orthogonal subcarriers and each subcarrier is allocated to at most one DL user and one UL user.
The FD BS is equipped with $N_{\mathrm{T}} > 1$  transmit antennas and a single receive antenna\footnote{Since there is no multiple access interference in the UL, the FD BS is equipped with a single receive antenna to reduce the hardware complexity.}. The $K+J+M$ users are single-antenna HD mobile communication devices to ensure low hardware complexity.
The DL and UL users are scheduled for simultaneous DL and UL transmission while idle users are not scheduled in the current time slot. However, the idle users may deliberately intercept the information signals intended for the DL and UL users. As a result, the idle users are treated as potential eavesdroppers which have to be taken into account for resource allocation algorithm design to guarantee communication security. In order to study the upper bound performance of the considered system, we assume that the FD BS has perfect channel state information (CSI) for resource allocation.

Assume that DL user $k$ and UL user $j$ are scheduled on subcarrier $i$ in a given scheduling time slot. The FD BS transmits a signal stream $\mathbf{w}_k^i d_k^{i \mathrm{DL}}$ to  DL user $k$ on subcarrier $i$, where $d_k^{i\mathrm{DL}}\in\mathbb{C}$ and $\mathbf{w}_k^i \in\mathbb{C}^{N_\mathrm{T}\times1}$ are the information bearing symbol for DL user $k$ and the corresponding beamforming vector on subcarrier $i$, respectively. Without loss of generality, we assume ${\cal E}\{\abs{d_k^{i\mathrm{DL}}}^2\}=1,\forall k\in\{1,\ldots,K\}$. Besides, in order to ensure secure communication, the FD BS transmits AN to interfere the reception of the idle users (potential eavesdroppers).
Therefore, the transmit signal vector on subcarrier $i$, $\mathbf{x}^i\in\mathbb{C}^{N_{\mathrm{T}}\times1}$, comprising data and AN, is given by $\mathbf{x}^i= \mathbf{w}_k^i d_k^{i \mathrm{DL}} +\mathbf{z}^i$, where $\mathbf{z}^i \in\mathbb{C}^{N_\mathrm{T}\times1}$ represents the AN vector on subcarrier $i$ generated by the FD BS to degrade the channel of the potential eavesdroppers on subcarrier $i$.
In particular, $\mathbf{z}^i$ is modeled as a complex Gaussian random vector with $\mathbf{z}^i\sim{\cal CN}(\mathbf{0},\mathbf{Z}^i)$,
where $\mathbf{Z}^i \in\mathbb{H}^{N_\mathrm{T}}$, $\mathbf{Z}^i \succeq \mathbf{0}$, denotes the covariance matrix of the AN. Therefore, the received signals at DL user $k\in\{1,\ldots,K\}$ and the FD BS on subcarrier $i$ are given by\vspace*{-2mm}
\begin{eqnarray}
\label{eqn:dl_user_rcv_signal}
\hspace*{-2mm} y_{k}^{i\mathrm{DL}}\hspace*{-3.5mm} &=&\hspace*{-3mm}\mathbf{h}_k^{iH}\mathbf{w}_k^i d_k^{i\mathrm{DL}} \hspace*{-0mm} + \hspace*{-0mm}\underbrace{\mathbf{h}_k^{iH} \mathbf{z}^i}_{\underset{\mbox{noise}}{\mbox{artificial}}}\hspace*{-0mm} + \hspace*{-0mm}\underbrace{ \sqrt{P_j^i}f_{j,k}^i d_j^{i\mathrm{UL}}}_{\underset{\mbox{interference}}{\mbox{co-channel}}}\hspace*{-0mm} + \hspace*{-0mm} n^{i\mathrm{DL}}_{k}, \\[-1mm]
\label{eqn:ul_rcv_signal}
\hspace*{-2mm} y^{i\mathrm{UL}}\hspace*{-3.5mm}&=&\hspace*{-3mm} \sqrt{P_j^i} g_j^i d_j^{i\mathrm{UL}}\hspace*{-0mm} + \hspace*{-3mm} \underbrace{\mathbf{h}_{\mathrm{SI}}^{iH}  \mathbf{w}_k^i d_k^{i\mathrm{DL}}}_{\mbox{self-interference}} \hspace*{-1mm} + \hspace*{-1mm} \underbrace{\mathbf{h}_{\mathrm{SI}}^{iH} \mathbf{z}^i}_{\underset{\mbox{noise}}{\mbox{artificial}}} \hspace*{-1mm} + \hspace*{0mm}n^{i\mathrm{UL}},
\end{eqnarray}
respectively. The channels between the FD BS and DL user $k$ and between UL user $j$ and DL user $k$ on subcarrier $i$ are denoted by $\mathbf{h}_k^i\in\mathbb{C}^{N_{\mathrm{T}}\times1}$ and $f_{j,k}^i \in\mathbb{C}$, respectively. $d_j^{i\mathrm{UL}}$, ${\cal E}\{\abs{d_j^{i\mathrm{UL}}}^2\}=1$, and $P_j^i$ denote the data symbol and transmit power of UL user $j$ on subcarrier $i$, respectively. $g_j^i\in\mathbb{C}$ denotes the channel between UL user $j$ and the FD BS on subcarrier $i$. Vector $\mathbf{h}_{\mathrm{SI}}^i \in{\mathbb{C}^{N_{\mathrm{T}}\times 1}}$ represents the self-interference (SI) channel of the FD BS on subcarrier $i$. Variables $\mathbf{h}_k^i$, $f_{j,k}^i$, $g_j^i$, and $\mathbf{h}_{\mathrm{SI}}^i$ capture the joint effect of path loss and small scale fading.  $n^{i\mathrm{UL}}\sim{\cal CN}(0,\sigma_{\mathrm{UL}}^2)$ and $n^{i\mathrm{DL}}_{k}\sim{\cal CN}(0,\sigma_{\mathrm{n}_k}^2)$ represent the additive white Gaussian noise (AWGN) at the FD BS and DL user $k$, respectively, where $\sigma_{\mathrm{UL}}^2$ and $\sigma_{\mathrm{n}_k}^2$ denote the corresponding noise powers, respectively.
In (\ref{eqn:dl_user_rcv_signal}), the term $\sqrt{P_j^i}f_{j,k}^i d_j^{i\mathrm{UL}}$ denotes the  co-channel interference (CCI) caused by UL user $j$ to DL user $k$ on subcarrier $i$. In (\ref{eqn:ul_rcv_signal}), the term $\mathbf{h}_{\mathrm{SI}}^{iH}  \mathbf{w}_k^i d_k^{i\mathrm{DL}}$ represents the SI.

Moreover, we assume the presence of $M$ potential eavesdroppers (idle users) and model them as a multiple-antenna HD device which is equipped with $M$ antennas. We note that one eavesdropper with $M$ antennas is equivalent to $M$ single-antenna eavesdroppers which are connected to a joint processing unit. The received signal at the equivalent multiple-antenna eavesdropper on subcarrier $i$ is given by \vspace*{-1mm}
\begin{eqnarray}
\label{eqn:eve_rcv_signal}\mathbf{y}^{i\mathrm{E}}\hspace*{-2mm} &=&\hspace*{-2mm} \mathbf{L}^{iH} \mathbf{w}_k^i d_k^{i\mathrm{DL}} \hspace*{0mm}+\hspace*{0mm} \sqrt{P_j^i}\mathbf{e}_{j}^i d_j^{i\mathrm{UL}}\hspace*{0mm} +\hspace*{-1mm} \underbrace{\mathbf{L}^{iH} \mathbf{z}^i}_{\mbox{artificial noise}}\hspace*{-1mm}  + \hspace*{0mm}\mathbf{n}^{i\mathrm{E}}.
\end{eqnarray}
Here, matrix $\mathbf{L}^i \in{\mathbb{C}^{N_{\mathrm{T}}\times M}}$ denotes the channel between the FD BS and the equivalent eavesdropper. Vector $\mathbf{e}_{j}^i\in\mathbb{C}^{M\times 1}$ denotes the channel between UL user $j$ and the equivalent eavesdropper on subcarrier $i$. $\mathbf{L}^i$ and $\mathbf{e}_{j}^i$ capture the joint effect of path loss and small scale fading. Finally, $\mathbf{n}^{i\mathrm{E}}\sim{\cal CN}(\mathbf{0},\sigma_{\mathrm{E}}^2\mathbf{I}_{M})$ represents the AWGN at the equivalent eavesdropper, where $\sigma_{\mathrm{E}}^2$ denotes the corresponding noise power.

\vspace*{-0mm}
\section{Resource Allocation Problem Formulation}
In this section, we formulate the resource allocation design as a non-convex optimization problem, after introducing the adopted performance metrics for the considered system. For the sake of notational simplicity, we define the following variables: $\mathbf{H}_k^i=\mathbf{h}_k^i\mathbf{h}_k^{iH}$, $k\in\{1,\ldots,K\}$, $\mathbf{H}_{\mathrm{SI}}^i=\mathbf{h}_{\mathrm{SI}}^i\mathbf{h}_{\mathrm{SI}}^{iH}$, $i\in\{1,\ldots,N_{\mathrm{F}}\}$.

\vspace*{-0mm}
\subsection{Weighted System Throughput and Secrecy Rate}
Assuming DL user $k$ and UL user $j$ are multiplexed on subcarrier $i$, the achievable rate (bits/s/Hz) of DL user $k$ and UL user $j$ on subcarrier $i$ are given by \vspace*{-2mm}
\begin{eqnarray} \label{rate-dl-k}
R_{k,j}^{i\mathrm{DL}}=&&\hspace*{-6mm} \log_2 \Big(  1  +  \frac{\abs{\mathbf{h}_{k}^{iH} \mathbf{w}_{k}^i }^2 } {\Tr(\mathbf{H}_k^i \mathbf{Z}^i) + P_j^i \abs{f_{j,k}^i}^2 + \sigma_{\mathrm{n}_k}^2}  \Big) \,\,\, \text{and}  \\[-1mm]
\label{rate-ul-j}
R_{k,j}^{i\mathrm{UL}}=&& \hspace*{-6mm} \log_2  \Big(1 + \frac{P_j^i\abs{g_j^i}^2}{\rho\big(\abs{\mathbf{h}_{\mathrm{SI}}^i \mathbf{w}_k^i}^2 + \Tr(\mathbf{H}_{\mathrm{SI}}^i \mathbf{Z}^i)\big) + \sigma_{\mathrm{UL}}^2}  \Big),
\end{eqnarray}
respectively. Therefore, the weighted system throughput on subcarrier $i$ is given by \vspace*{-2.3mm}
\begin{eqnarray} \label{throughput_i}
U_{k,j}^i(\mathbf{s},\mathbf{W},\mathbf{p},\mathbf{Z}) = s_{k,j}^{i}\Big[  w_k  R_{k,j}^{i\mathrm{DL}} +  \mu_j  R_{k,j}^{i\mathrm{UL}} \Big],
\end{eqnarray}
where $s_{k,j}^{i} \in\{0,1\}$ is the subcarrier allocation indicator. Specifically, $s_{k,j}^{i}=1$ if DL user $k$ and UL $j$ are multiplexed on subcarrier $i$ and $s_{m,n}^{i}=0$ if another resource allocation policy is used.
The positive constants $0  \le  w_k \le 1$  and $0  \le  \mu_j \le 1$ denote the priorities of DL user $k$ and UL user $j$ in resource allocation, respectively, and are specified in the media access control (MAC) layer to achieve certain fairness objectives. $0 <\rho \ll 1$ is a constant modelling the noisiness of the SI cancellation at the FD BS. To facilitate the presentation, we introduce $\mathbf{s}\in\mathbb{Z}^{{N_{\mathrm{F}}}K^2 \times 1}$, $\mathbf{W}\in\mathbb{C}^{{N_{\mathrm{F}}}K \times N_{\mathrm{T}}}$, $\mathbf{p}\in\mathbb{R}^{{N_{\mathrm{F}}}J \times 1}$, and $\mathbf{Z}\in\mathbb{C}^{{N_{\mathrm{F}}} N_{\mathrm{T}} \times M}$ as the collections of the optimization variables $s_{k,j}^i, \forall i,k,j$, $\mathbf{w}_k^i, \forall i,k$, $P_j^i, \forall i,j$, and $\mathbf{Z}^i, \forall i$, respectively.

Next, for guaranteeing communication security in the considered system, we design the resource allocation algorithm under a worst-case assumption. In particular, we assume that the equivalent eavesdropper can cancel the UL (DL) user's interference before decoding the information of the desired DL (UL) user on each subcarrier.
Thus, under this assumption, the capacity of the channel of DL user $k$ and UL user $j$ on subcarrier $i$ with respect to the equivalent eavesdropper can be written as\vspace*{-0mm}
\begin{eqnarray}
\label{eqn:DL-EVE-rate}C_{k}^{i\mathrm{DL-E}}\hspace*{-1.5mm}&=&\hspace*{-1.5mm} \log_2\det(\mathbf{I}_{N_{\mathrm{E}}}+ (\mathbf{X}^{i})^{-1}\mathbf{L}^{iH} \mathbf{w}_k^i \mathbf{w}_k^{iH} \mathbf{L}^i) \,\, \mbox{and}\\[-1mm]
\label{eqn:UL-EVE-rate}C_{j}^{i\mathrm{UL-E}}\hspace*{-1.5mm}&=&\hspace*{-1.5mm} \log_2\det(\mathbf{I}_{N_{\mathrm{E}}}+ P_j^i (\mathbf{X}^{i})^{-1}\mathbf{e}_{j}^i\mathbf{e}_{j}^{iH}),
\end{eqnarray}
respectively, where $\mathbf{X}^i=\mathbf{L}^{iH} \mathbf{Z}^i \mathbf{L}^i +\sigma_{\mathrm{E}}^2\mathbf{I}_{N_\mathrm{E}}$ denotes the interference-plus-noise covariance matrix of the equivalent eavesdropper on subcarrier $i$. The achievable secrecy rates between the FD BS and DL user $k$ and UL user $j$ on subcarrier $i$ are given by $R_{k,j}^{i\mathrm{DL-Sec}}\hspace*{-1.5mm}=\hspace*{-1.5mm}\Big[R_{k,j}^{i\mathrm{DL}}- C_{k}^{i\mathrm{DL-E}}\Big]^+$ and $R_{k,j}^{i\mathrm{UL-Sec}}\hspace*{-1.5mm}=\hspace*{-1.5mm}\Big[R_{k,j}^{i\mathrm{UL}}- C_{j}^{i\mathrm{UL-E}}\Big]^+$,
respectively.

\vspace*{-0mm}
\subsection{Optimization Problem Formulation}
The system design objective is the maximization of the weighted system throughput. The resource allocation policy is obtained by solving the following optimization problem:\vspace*{-2mm}
\begin{eqnarray} \label{pro}
&&\hspace*{-7mm}\underset{\mathbf{s},\mathbf{W},\mathbf{p},\mathbf{Z}}{\maxo}\,\, \,\,  \sum_{i=1}^{N_{\mathrm{F}}}\sum_{k=1}^{K} \sum_{j=1}^{J} U_{k,j}^i(\mathbf{s},\mathbf{W},\mathbf{p},\mathbf{Z}) \notag\\  [-1mm]
\notag\mbox{s.t.}
&&\hspace*{-5mm}\mbox{C1: }\overset{N_{\mathrm{F}}}{\underset{i=1}{\sum}} \overset{K}{\underset{k=1}{\sum}} \overset{J}{\underset{j=1}{\sum}} s_{k,j}^i (\norm{\mathbf{w}_{k}^i}^2 + \Tr(\mathbf{Z}^i)) \le P_{\mathrm{max}}^{\mathrm{DL}}, \notag\\ [-1mm]
&&\hspace*{-5mm}\mbox{C2: }\overset{N_{\mathrm{F}}}{\underset{i=1}{\sum}} \overset{K}{\underset{k=1}{\sum}} s_{k,j}^i P_j^i \le P_{\mathrm{max}_j}^{\mathrm{UL}}, \forall j,\notag
\quad \mbox{C3: } P_j^i \ge 0, \forall i,j,\\ [-1mm]
&&\hspace*{-5mm}\mbox{C4: }s_{k,j}^i C_{k}^{i\mathrm{DL-E}} \hspace*{-0.7mm} \le \hspace*{-0.7mm} R_{\mathrm{tol}_{k}}^{i\mathrm{DL}},  \,\,\, \notag
 \mbox{C5: }s_{k,j}^i  C_{j}^{i\mathrm{UL-E}} \hspace*{-0.7mm}\le \hspace*{-0.7mm} R_{\mathrm{tol}_{j}}^{i\mathrm{UL}},  \\[-1mm]
&&\hspace*{-5mm}\mbox{C6: } s_{k,j}^i \in \{0,1\}, \forall i,k,j, \,\,\,\,\, \mbox{C7: } \overset{K}{\underset{k=1}{\sum}} \overset{J}{\underset{j=1}{\sum}} s_{k,j}^i \le 1, \forall i, \notag\\ [-1mm]
&&\hspace*{-5mm}\mbox{C8: } \mathbf{Z}^i \succeq \mathbf{0}, \,\mathbf{Z}^i \in\mathbb{H}^{N_\mathrm{T}}, \, \forall i.
\end{eqnarray}
Constraint C1 is the power constraint for the BS with maximum transmit power allowance $P_{\mathrm{max}}^{\mathrm{DL}}$. Constraint C2 limits the transmit power of UL user $j$ to $ P_{\mathrm{max}_j}^{\mathrm{UL}}$. Constraint C3 ensures that the power of UL user $j$ is non-negative. $R_{\mathrm{tol}_{k}}^{i\mathrm{DL}}$ and $R_{\mathrm{tol}_{j}}^{i\mathrm{UL}}$, in C4 and C5, respectively, are pre-defined system parameters representing the maximum tolerable data rate at the potential eavesdropper for decoding the information of DL user $k$ and UL user $j$ on subcarrier $i$, respectively.
If the above optimization problem is feasible, the proposed problem formulation guarantees that the secrecy rate for DL user $k$ is bounded below as $R_{k}^{\mathrm{DL-Sec}}\hspace*{-1mm}\ge\hspace*{-1mm}\sum_{i=1}^{N_{\mathrm{F}}}\sum_{j=1}^J \hspace*{-0.5mm}s_{k,j}^i\hspace*{-0.5mm}\Big(R_{k,j}^{i\mathrm{DL}}- R_{\mathrm{tol}_{k}}^{i\mathrm{DL}}\Big)$
and the secrecy rate for UL user $j$ is bounded below as
$R_{j}^{\mathrm{UL-Sec}}\hspace*{-1mm}\ge\hspace*{-1mm}\sum_{i=1}^{N_{\mathrm{F}}}\sum_{k=1}^K \hspace*{-0.5mm}s_{k,j}^i \hspace*{-0.5mm}\Big(R_{k,j}^{i\mathrm{UL}}- R_{\mathrm{tol}_{j}}^{i\mathrm{UL}}\Big)$.
Constraints C6 and C7 are imposed to guarantee that each subcarrier is allocated to at most one DL user and one UL user.  Constraint C8 is imposed since covariance matrix $\mathbf{Z}^i$ has to be a Hermitian positive semidefinite matrix.

The considered resource allocation optimization problem in \eqref{pro} is a mixed combinatorial non-convex optimization problem, and obtaining the globally optimal solution entails a prohibitively high computational complexity.  Therefore, in the next section, we propose an efficient suboptimal scheme based on successive convex approximation \cite{sun2016optimalJournal}.

\vspace*{-0mm}
\section{Solution of the Optimization Problem}
In this section, we propose a suboptimal algorithm with low computational complexity\footnote{The proposed algorithm has a polynomial time complexity which is desirable for real-time
implementation \cite[Chapter 34]{book:polynoimal}.},  which finds a locally optimal solution for the optimization problem in \eqref{pro}.

Let us define $\mathbf{W}_{k}^i=\mathbf{w}_k^i \mathbf{w}_k^{iH}$, $\mathbf{W}_{k}^i  \in \mathbb{H}^{N_\mathrm{T}}$.
Then, we rewrite the weighted system throughput of DL user $k$ and UL user $j$ on subcarrier $i$ in \eqref{throughput_i} as:\vspace*{-2mm}
\begin{eqnarray} \label{sdp-throughput_i}
&&U_{k,j}^i(\mathbf{s},\mathbf{W},\mathbf{p},\mathbf{Z}) \\[-1mm]
\hspace*{-2mm}&=&\hspace*{-2mm} w_k  \log_2 \Big(  1  +  \frac{s_{k,j}^{i} \Tr(\mathbf{H}_k^i \mathbf{W}_k^i) } {s_{k,j}^{i}\Tr(\mathbf{H}_k^i \mathbf{Z}^i) + s_{k,j}^{i} P_j^i \abs{f_{j,k}^i}^2 + \sigma_{\mathrm{n}_k}^2}  \Big) \notag \\[-1mm]
\hspace*{-2mm}&+&\hspace*{-2mm}  \mu_j  \log_2  \Big(1 + \frac{s_{k,j}^{i} P_j^i\abs{g_j^i}^2}{\rho s_{k,j}^{i} \Tr\big(\mathbf{H}_{\mathrm{SI}}^i (\mathbf{W}_k^i+\mathbf{Z}^i)\big) + \sigma_{\mathrm{UL}}^2}  \Big) . \notag
\end{eqnarray}
The product terms between $s_{k,j}^{i}$ and other optimization variables in \eqref{sdp-throughput_i}, i.e., $s_{k,j}^{i} \Tr(\mathbf{H}_k^i \mathbf{W}_k^i)$, $s_{k,j}^{i} P_j^i$, and $s_{k,j}^{i} \Tr(\mathbf{H}_k^i \mathbf{Z}^i)$, are obstacles in the design of a computationally efficient resource allocation algorithm. Hence, we employ the big-M method to overcome this difficulty \cite{lee2011mixed}. In particular, we first define $\tilde{\mathbf{W}}_{k,j}^i=s_{k,j}^{i} \mathbf{W}_k^i$, $\tilde{\mathbf{W}}_{k,j}^i  \in \mathbb{H}^{N_\mathrm{T}}$, $\tilde{\mathbf{Z}}_{k,j}^i=s_{k,j}^{i} \mathbf{Z}^i$, $\tilde{\mathbf{Z}}_{k,j}^i  \in \mathbb{H}^{N_\mathrm{T}}$, and $\tilde{P}_{k,j}^i=s_{k,j}^{i} P_j^i$, and then rewrite the weighted system throughput in \eqref{sdp-throughput_i} as: \vspace*{-2mm}
\begin{eqnarray} \label{big-M-throughput_i}
&&U_{k,j}^i(\tilde{\mathbf{W}},\tilde{\mathbf{p}},\tilde{\mathbf{Z}}) \\[-3mm]
\hspace*{-2mm}&=&\hspace*{-2mm} w_k  \log_2 \Big(  1  +  \frac{ \Tr(\mathbf{H}_k^i \tilde{\mathbf{W}}_{k,j}^i) } {\Tr(\mathbf{H}_k^i \tilde{\mathbf{Z}}_{k,j}^i) + \tilde{P}_{k,j}^i \abs{f_{j,k}^i}^2 + \sigma_{\mathrm{n}_k}^2}  \Big) \notag \\[-1mm]
\hspace*{-2mm}&+&\hspace*{-2mm}  \mu_j  \log_2  \Big(1 + \frac{\tilde{P}_{k,j}^i\abs{g_j^i}^2}{\rho\Tr\big(\mathbf{H}_{\mathrm{SI}}^i (\tilde{\mathbf{W}}_{k,j}^i+\tilde{\mathbf{Z}}_{k,j}^i)\big) + \sigma_{\mathrm{UL}}^2}  \Big) , \notag
\end{eqnarray}
where $\tilde{\mathbf{W}}$, $\tilde{\mathbf{p}}$, and $\tilde{\mathbf{Z}}$ are the collections of all $\tilde{\mathbf{W}}_{k,j}^i$, $\tilde{P}_{k,j}^i$, and $\tilde{\mathbf{Z}}_{k,j}^i$, respectively. Next, we decompose the product terms by imposing the following additional constraints:\vspace*{-2mm}
\begin{eqnarray}
\hspace*{-9mm}&&\hspace*{-4mm}\mbox{C9: } \hspace*{-1mm} \tilde{\mathbf{W}}_{k,j}^i \hspace*{-1mm} \preceq \hspace*{-1mm}  P_{\mathrm{max}}^{\mathrm{DL}} \mathbf{I}_{N_{\mathrm{T}}} s_{k,j}^i,
\quad \mbox{C10: } \hspace*{-1mm} \tilde{\mathbf{W}}_{k,j}^i \hspace*{-1mm} \preceq \hspace*{-1mm} \mathbf{W}_k^i,
\\
\hspace*{-9mm}&&\hspace*{-4mm}\mbox{C11: } \hspace*{-1mm} \tilde{\mathbf{W}}_{k,j}^i \hspace*{-1mm} \succeq \hspace*{-1mm} \mathbf{W}_k^i \hspace*{-1mm} - \hspace*{-1mm}(1 \hspace*{-1mm} - \hspace*{-1mm}s_{k,j}^i)P_{\mathrm{max}}^{\mathrm{DL}}\mathbf{I}_{N_{\mathrm{T}}}, \quad  \mbox{C12: }  \hspace*{-1mm} \tilde{\mathbf{W}}_{k,j}^i \hspace*{-1mm} \succeq \hspace*{-1mm} \mathbf{0}, \\
\hspace*{-9mm}&&\hspace*{-4mm}\mbox{C13: } \hspace*{-1mm} \tilde{\mathbf{Z}}_{k,j}^i \hspace*{-1mm} \preceq \hspace*{-1mm}  P_{\mathrm{max}}^{\mathrm{DL}} \mathbf{I}_{N_{\mathrm{T}}} s_{k,j}^i,
\quad \mbox{C14: } \hspace*{-1mm} \tilde{\mathbf{Z}}_{k,j}^i \hspace*{-1mm} \preceq \hspace*{-1mm} \mathbf{Z}^i,
\\
\hspace*{-9mm}&&\hspace*{-4mm}\mbox{C15: } \hspace*{-1mm} \tilde{\mathbf{Z}}_{k,j}^i \hspace*{-1mm} \succeq \hspace*{-1mm} \mathbf{Z}^i \hspace*{-1mm} - \hspace*{-1mm}(1 \hspace*{-1mm} - \hspace*{-1mm}s_{k,j}^i)P_{\mathrm{max}}^{\mathrm{DL}}\mathbf{I}_{N_{\mathrm{T}}}, \quad  \mbox{C16: }  \hspace*{-1mm} \tilde{\mathbf{Z}}_{k,j}^i \hspace*{-1mm} \succeq \hspace*{-1mm} \mathbf{0},\\
\hspace*{-9mm}&&\hspace*{-4mm}\mbox{C17: } \hspace*{-1mm} \tilde{P}_{k,j}^i \hspace*{-1mm} \le \hspace*{-1mm}  P_{\mathrm{max}_j}^{\mathrm{UL}} s_{k,j}^i,
\quad \mbox{C18: } \hspace*{-1mm} \tilde{P}_{k,j}^i \hspace*{-1mm} \le \hspace*{-1mm} P_j^i,
\\
\hspace*{-9mm}&&\hspace*{-4mm}\mbox{C19: } \hspace*{-1mm} \tilde{P}_{k,j}^i \hspace*{-1mm} \ge \hspace*{-1mm} P_j^i \hspace*{-1mm} - \hspace*{-1mm}(1 \hspace*{-1mm} - \hspace*{-1mm}s_{k,j}^i)P_{\mathrm{max}_j}^{\mathrm{UL}}, \quad  \mbox{C20: }  \hspace*{-1mm} \tilde{P}_{k,j}^i \hspace*{-1mm} \ge \hspace*{-1mm} 0.
\end{eqnarray}

With the aforementioned definitions, we rewrite constraints C4 and C5 as: \vspace*{-2mm}
\begin{eqnarray} \label{big-M-C4}
&&\hspace*{-10mm} \mbox{C4: }\log_2\det(\mathbf{I}_{N_{\mathrm{E}}}+ (\tilde{\mathbf{X}}_{k,j}^{i})^{-1}\mathbf{L}^{iH} \tilde{\mathbf{W}}_{k,j}^i  \mathbf{L}^i) \hspace*{-1mm} \le \hspace*{-1mm} R_{\mathrm{tol}_{k}}^{i\mathrm{DL}}, \, \forall i,k,j,\\[-1mm]
\label{big-M-C5}
&&\hspace*{-10mm} \mbox{C5: } \log_2\det(\mathbf{I}_{N_{\mathrm{E}}}+ \tilde{P}_{k,j}^i (\tilde{\mathbf{X}}_{k,j}^{i})^{-1}\mathbf{e}_{j}^i\mathbf{e}_{j}^{iH}) \hspace*{-1mm} \le \hspace*{-1mm} R_{\mathrm{tol}_{j}}^{i\mathrm{UL}}, \, \forall i,k,j,
\end{eqnarray}
respectively, where $\tilde{\mathbf{X}}_{k,j}^{i} = \mathbf{L}^{iH} \tilde{\mathbf{Z}}_{k,j}^i \mathbf{L}^i +\sigma_{\mathrm{E}}^2\mathbf{I}_{N_\mathrm{E}}$.
Now, the original optimization problem in \eqref{pro} can be rewritten in the following equivalent form: \vspace*{-2mm}
\begin{eqnarray} \label{sdp-pro}
&&\hspace*{-7mm}\underset{\mathbf{s},\tilde{\mathbf{W}},\tilde{\mathbf{p}},\tilde{\mathbf{Z}}}{\maxo}\,\, \,\,  \sum_{i=1}^{N_{\mathrm{F}}}\sum_{k=1}^{K} \sum_{j=1}^{J} U_{k,j}^i(\tilde{\mathbf{W}},\tilde{\mathbf{p}},\tilde{\mathbf{Z}})\\  [-1mm]
\notag\mbox{s.t.}
&&\hspace*{-6mm}\mbox{C1: }\overset{N_{\mathrm{F}}}{\underset{i=1}{\sum}} \overset{K}{\underset{k=1}{\sum}} \overset{J}{\underset{j=1}{\sum}}  \Tr(\tilde{\mathbf{W}}_{k,j}^i) + \Tr(\tilde{\mathbf{Z}}_{k,j}^i) \le P_{\mathrm{max}}^{\mathrm{DL}},   \notag\\ [-1mm]
&&\hspace*{-6mm}\mbox{C2: }\overset{N_{\mathrm{F}}}{\underset{i=1}{\sum}} \overset{K}{\underset{k=1}{\sum}} \tilde{P}_{k,j}^i \le P_{\mathrm{max}_j}^{\mathrm{UL}}, \forall j, \quad \mbox{C3--C20},\notag \\[-1mm]
&&\hspace*{-6mm}\mbox{C21}\mbox{: }\tilde{\mathbf{W}}_{k,j}^i \hspace*{-0.5mm} \succeq \hspace*{-0.5mm} \mathbf{0}, \forall i,k,j,\,\, \mbox{C22:}\Rank(\tilde{\mathbf{W}}_{k,j}^i) \hspace*{-0.5mm} \le \hspace*{-0.5mm} 1, \forall i,k,j, \notag
\end{eqnarray}
where constraints C21 and C22 are imposed to guarantee that $\tilde{\mathbf{W}}_{k,j}^i=s_{k,j}^i\mathbf{w}_k^i\mathbf{w}_k^{iH}$ holds after optimization.

In problem \eqref{sdp-pro}, constraints C4 and C5 are non-convex constraints. Hence, we establish the following proposition to facilitate the transformation of these constraints.

\begin{Prop}
For $R_{\mathrm{tol}_{k}}^{i\mathrm{DL}} > 0$ and $R_{\mathrm{tol}_{j}}^{i\mathrm{UL}} > 0$, we have the following implications for constraints ${\mbox{C4}}$ and ${\mbox{C5}}$ of problem \eqref{sdp-pro}, respectively: \vspace*{-2mm}
\begin{eqnarray}
\label{eqn:DL-tol-rate-QC}{\mbox{C4}}\hspace*{-2mm}&\Rightarrow&\hspace*{-2mm} {\widetilde{\mbox{C4}}}\mbox{: } \mathbf{L}^{iH}\tilde{\mathbf{W}}_{k,j}^i \mathbf{L}^i \hspace*{1mm}\preceq\hspace*{1mm} \xi_{k}^{i\mathrm{DL}}\tilde{\mathbf{X}}_{k,j}^{i}, \, \forall i,k,j, \,\,\,\, \text{and}\\
\label{eqn:UL-tol-rate-QC}{\mbox{C5}}\hspace*{-2mm}&\Leftrightarrow&\hspace*{-2mm} {\widetilde{\mbox{C5}}}\mbox{: } \tilde{P}_{k,j}^i \mathbf{e}_{j}^i \mathbf{e}_{j}^{iH} \hspace*{1mm}\preceq\hspace*{1mm} \xi_{j}^{i\mathrm{UL}} \tilde{\mathbf{X}}_{k,j}^{i}, \, \forall i,k,j,
\end{eqnarray}
where $\xi_{k}^{i\mathrm{DL}}=2^{R_{\mathrm{tol}_{k}}^{i\mathrm{DL}}}-1$ and $\xi_{j}^{i\mathrm{UL}}=2^{R_{\mathrm{tol}_{j}}^{i\mathrm{UL}}}-1$. We note that ${\mbox{C4}}$ and ${\widetilde{\mbox{C4}}}$ are equivalent if $\Rank(\tilde{\mathbf{W}}_{k,j}^i) \le 1$. Besides, ${\mbox{C5}}$ and ${\widetilde{\mbox{C5}}}$ are always equivalent.
\end{Prop}
\emph{\quad Proof: } The proof can be found in Appendix-A in \cite{Sun16FDSecurity}. \hfill\qed

We note that the resulting constraints ${\widetilde{\mbox{C4}}}$ and ${\widetilde{\mbox{C5}}}$ are convex constraints. Besides, in order to handle the non-convex integer constraint C6 in problem \eqref{sdp-pro}, we rewrite constraint C6 in equivalent form: \vspace*{-2mm}
\begin{eqnarray}
\hspace*{-0mm}\text{C6}\mbox{a: } \hspace*{-1mm} \overset{{N_{\mathrm{F}}}}{\underset{i=1}{\sum}}  \overset{K}{\underset{k=1}{\sum}}  \overset{J}{\underset{j=1}{\sum}}  s_{k,j}^i \hspace*{-0.5mm} - \hspace*{-0.5mm} (s_{k,j}^i)^2 \hspace*{-0.5mm} \le \hspace*{-0.5mm}  0 \,\, \text{and}
\text{ C6}\mbox{b: } 0 \hspace*{-0.5mm} \le \hspace*{-0.5mm} s_{k,j}^i \hspace*{-0.5mm} \le \hspace*{-0.5mm} 1,
\end{eqnarray}
i.e., optimization variables $s_{k,j}^i$ are relaxed to a continuous interval between zero and one. However, constraint $\mbox{C6a}$ is a reverse convex function \cite{ng2015power} which makes problem \eqref{sdp-pro} still non-convex.
To resolve this issue, we reformulate problem \eqref{sdp-pro} as \vspace*{-3mm}
\begin{eqnarray} \label{penalty-pro}
&&\hspace*{-7mm}\underset{\mathbf{s},\tilde{\mathbf{W}},\tilde{\mathbf{p}},\tilde{\mathbf{Z}}}{\mino}\,\, \,\,  \sum_{i=1}^{N_{\mathrm{F}}}\sum_{k=1}^{K} \sum_{j=1}^{J} -U_{k,j}^i(\tilde{\mathbf{W}},\tilde{\mathbf{p}},\tilde{\mathbf{Z}}) +  \eta\big(s_{k,j}^i -(s_{k,j}^i)^2\big)  \notag \\  [-0mm]
&&\mbox{s.t.} \hspace*{5mm}\mbox{C1--C3},\widetilde{\mbox{C4}},\widetilde{\mbox{C5}},\mbox{C6b}, \mbox{C7-C22},
\end{eqnarray}
where $\eta \gg 1$ acts as a penalty factor for penalizing the objective function for any $s_{k,j}^i$ that is not equal to $0$ or $1$. It is shown in \cite{sun2016optimalJournal,ng2015power} that \eqref{penalty-pro} and \eqref{sdp-pro} are equivalent for $\eta \gg 1$.

\begin{table} \vspace*{-3mm}
\begin{algorithm} [H]                    
\caption{Successive Convex Approximation}          
\label{alg1}                           
\begin{algorithmic} [1]
\small          
\STATE Initialize the maximum number of iterations $I_{\mathrm{max}}$, penalty factor $\eta \hspace*{-0.8mm} \gg \hspace*{-0.8mm} 1$, iteration index $m\hspace*{-0.8mm} =\hspace*{-0.8mm}1$, and initial point $\mathbf{s}^{(1)}$, $\tilde{\mathbf{W}}^{(1)}$, $\tilde{\mathbf{Z}}^{(1)}$, and $\tilde{\mathbf{p}}^{(1)}$

\REPEAT
\STATE Solve \eqref{sdr-dc} for a given $\mathbf{s}^{(m)}$, $\tilde{\mathbf{W}}^{(m)}$, $\tilde{\mathbf{Z}}^{(m)}$, and $\tilde{\mathbf{p}}^{(m)}$ and store the intermediate resource allocation policy $\{\mathbf{s}, \tilde{\mathbf{W}}, \tilde{\mathbf{Z}},\tilde{\mathbf{p}}\}$

\STATE Set $m=m+1$ and $\mathbf{s}^{(m)}=\mathbf{s}$ , $\tilde{\mathbf{W}}^{(m)}=\tilde{\mathbf{W}}$, $\tilde{\mathbf{Z}}^{(m)}=\tilde{\mathbf{Z}}$, and $\tilde{\mathbf{p}}^{(m)}=\tilde{\mathbf{p}}$

\UNTIL convergence or $k=I_{\mathrm{max}}$

\STATE $\mathbf{s}^{*}=\mathbf{s}^{(m)}$, $\tilde{\mathbf{W}}^{*}=\tilde{\mathbf{W}}^{(m)}$,  $\tilde{\mathbf{Z}}^{*}=\tilde{\mathbf{Z}}^{(m)}$, and $\tilde{\mathbf{p}}^{*}=\tilde{\mathbf{p}}^{(m)}$

\end{algorithmic}
\end{algorithm}\vspace*{-10mm}
\end{table}

The resulting optimization problem in \eqref{penalty-pro} is still non-convex because of the objective function. To facilitate the presentation, we rewrite problem \eqref{penalty-pro} as \vspace*{-2mm}
\begin{eqnarray} \label{dc-penalty-pro}
&&\hspace*{-7mm}\underset{\mathbf{s},\tilde{\mathbf{W}},\tilde{\mathbf{p}},\tilde{\mathbf{Z}}}{\mino}\,\, \,\,   F(\tilde{\mathbf{W}},\tilde{\mathbf{p}},\tilde{\mathbf{Z}})-G(\tilde{\mathbf{W}},\tilde{\mathbf{p}},\tilde{\mathbf{Z}}) \notag  +  \eta\big(H(\mathbf{s})-M(\mathbf{s})\big)  \notag \\  [-0mm]
&&\mbox{s.t.} \hspace*{5mm}\mbox{C1--C3},\widetilde{\mbox{C4}},\widetilde{\mbox{C5}},\mbox{C6b}, \mbox{C7-C22},
\end{eqnarray} \vspace*{-2mm}
where
\begin{eqnarray}
\hspace*{-4mm}&&F(\tilde{\mathbf{W}},\tilde{\mathbf{p}},\tilde{\mathbf{Z}}) \notag \\ [-0mm]
\hspace*{-4mm}&=&\hspace*{-2mm} \overset{{N_{\mathrm{F}}}}{\underset{i=1}{\sum}}  \overset{K}{\underset{k=1}{\sum}}  \overset{J}{\underset{j=1}{\sum}} w_k  \log_2 \hspace*{-1mm} \Big(\hspace*{-1mm}\Tr \hspace*{-1mm} \big(\mathbf{H}_k^i (\tilde{\mathbf{W}}_{k,j}^i \hspace*{-1mm} + \hspace*{-1mm} \tilde{\mathbf{Z}}_{k,j}^i)\big) \hspace*{-1mm} + \hspace*{-1mm} \tilde{P}_{k,j}^i \abs{f_{j,k}^i}^2 \hspace*{-1mm} + \hspace*{-1mm} \sigma_{\mathrm{n}_k}^2 \Big) \notag \\[-0mm]
\hspace*{-4mm}&+&\hspace*{-2mm} \mu_j  \log_2 \hspace*{-0.8mm} \Big(\rho\Tr \hspace*{-0.8mm} \big(\mathbf{H}_{\mathrm{SI}}^i (\tilde{\mathbf{W}}_{k,j}^i \hspace*{-0.8mm} + \hspace*{-0.8mm} \tilde{\mathbf{Z}}_{k,j}^i)\big) \hspace*{-0.8mm}  + \hspace*{-0.8mm} \tilde{P}_{k,j}^i\abs{g_j^i}^2 + \hspace*{-0.8mm} \sigma_{\mathrm{UL}}^2  \Big),\\[-0mm]
\hspace*{-4mm}&&G(\tilde{\mathbf{W}},\tilde{\mathbf{p}},\tilde{\mathbf{Z}}) \notag\\[-1mm]
\hspace*{-4mm}&=&\hspace*{-2mm} \overset{{N_{\mathrm{F}}}}{\underset{i=1}{\sum}}  \overset{K}{\underset{k=1}{\sum}}  \overset{J}{\underset{j=1}{\sum}} w_k \log_2\Big(\Tr(\mathbf{H}_k^i \tilde{\mathbf{Z}}_{k,j}^i) + \tilde{P}_{k,j}^i \abs{f_{j,k}^i}^2 + \sigma_{\mathrm{n}_k}^2 \Big) \notag \\[-0mm]
\hspace*{-4mm}&+&\hspace*{-2mm} \mu_j \log_2 \Big( \rho\Tr\big(\mathbf{H}_{\mathrm{SI}}^i (\tilde{\mathbf{W}}_{k,j}^i+\tilde{\mathbf{Z}}_{k,j}^i)\big) + \sigma_{\mathrm{UL}}^2 \Big), \,\,\, \\[-0mm]
\hspace*{-4mm}&& \hspace*{-6mm}H(\mathbf{s}) \hspace*{-1mm}=\hspace*{-1mm} \overset{{N_{\mathrm{F}}}}{\underset{i=1}{\sum}} \overset{K}{\underset{k=1}{\sum}} \overset{J}{\underset{j=1}{\sum}}s_{k,j}^i,  \,\, \text{and}\,\,
M(\mathbf{s}) \hspace*{-1mm}=\hspace*{-1mm} \overset{{N_{\mathrm{F}}}}{\underset{i=1}{\sum}} \overset{K}{\underset{k=1}{\sum}} \overset{J}{\underset{j=1}{\sum}}(s_{k,j}^i)^2.
\end{eqnarray}

We note that problem \eqref{dc-penalty-pro} is in the canonical form of difference of convex (d.c.) function programs. Therefore, we can obtain a locally optimal solution of \eqref{dc-penalty-pro} by applying successive convex approximation \cite{dinh2010local}.
In particular, since $G(\tilde{\mathbf{W}},\tilde{\mathbf{p}},\tilde{\mathbf{Z}})$ is a differentiable convex function, for any feasible point $\tilde{\mathbf{W}}^{(m)}$, $\tilde{\mathbf{p}}^{(m)}$, and $\tilde{\mathbf{Z}}^{(m)}$ we have the following inequality: \vspace*{-2mm}
\begin{eqnarray}\label{ineq1}
&& \hspace*{-6mm}G (\tilde{\mathbf{W}},\tilde{\mathbf{p}},\tilde{\mathbf{Z}}) \ge  G  (\tilde{\mathbf{W}}^{(m)},\tilde{\mathbf{p}}^{(m)},\tilde{\mathbf{Z}}^{(m)}) \notag \\[-0mm]
\hspace*{-2mm}&+& \hspace*{-2mm}  \Tr(\nabla_{\tilde{\mathbf{W}}} G (\tilde{\mathbf{W}}^{(m)},\tilde{\mathbf{p}}^{(m)},\tilde{\mathbf{Z}}^{(m)})^{T}(\tilde{\mathbf{W}} \hspace*{-1mm} - \hspace*{-1mm}\tilde{\mathbf{W}}^{(m)}) ) \notag \\[-0mm]
\hspace*{-2mm}&+& \hspace*{-2mm}  \Tr(\nabla_{\tilde{\mathbf{p}}} G (\tilde{\mathbf{W}}^{(m)},\tilde{\mathbf{p}}^{(m)},\tilde{\mathbf{Z}}^{(m)})^{T}(\tilde{\mathbf{p}} \hspace*{-1mm} - \hspace*{-1mm}\tilde{\mathbf{p}}^{(m)}) ) \notag \\[-0mm]
\hspace*{-2mm}&+& \hspace*{-2mm}  \Tr(\nabla_{\tilde{\mathbf{Z}}} G (\tilde{\mathbf{W}}^{(m)},\tilde{\mathbf{p}}^{(m)},\tilde{\mathbf{Z}}^{(m)})^{T}(\tilde{\mathbf{Z}} \hspace*{-1mm} - \hspace*{-1mm}\tilde{\mathbf{Z}}^{(m)}) ) \notag \\[-0mm]
\hspace*{-2mm} &\triangleq& \hspace*{-2mm} \overline{G} (\tilde{\mathbf{W}},\tilde{\mathbf{p}},\tilde{\mathbf{Z}},\tilde{\mathbf{W}}^{(m)},\tilde{\mathbf{p}}^{(m)},\tilde{\mathbf{Z}}^{(m)}),
\end{eqnarray}
where the right hand side of \eqref{ineq1} is an affine function and represents the global underestimation of $G (\tilde{\mathbf{W}},\tilde{\mathbf{p}},\tilde{\mathbf{Z}})$. Similarly, we denote $\overline{M} (\mathbf{s},  \mathbf{s}^{(m)})$ as the global underestimation of $M(\mathbf{s})$. Besides, the non-convexity of problem \eqref{dc-penalty-pro} also comes from the rank-one constraint C22. Using a similar approach as in \cite{Sun16FDSecurity}, we apply semidefinite programming (SDP) relaxation by removing constraint C22.
Therefore, for any given $\mathbf{s}^{(m)}$, $\tilde{\mathbf{W}}^{(m)}$, $\tilde{\mathbf{Z}}^{(m)}$, and $\tilde{\mathbf{p}}^{(m)}$, we can obtain a lower bound of \eqref{dc-penalty-pro} by solving the following optimization problem: \vspace*{-1mm}
\begin{eqnarray}\label{sdr-dc}
\hspace*{-1mm}&&\hspace*{-7mm}
\underset{\mathbf{s},\tilde{\mathbf{W}},\tilde{\mathbf{p}},\tilde{\mathbf{Z}}}{\mino} \,\, F(\tilde{\mathbf{W}},\tilde{\mathbf{p}},\tilde{\mathbf{Z}})-\overline{G} (\tilde{\mathbf{W}},\tilde{\mathbf{p}},\tilde{\mathbf{Z}},\tilde{\mathbf{W}}^{(m)},\tilde{\mathbf{p}}^{(m)},\tilde{\mathbf{Z}}^{(m)}) \notag \\[-2mm]
\hspace*{-1mm}&&\hspace*{8mm} + \eta \big(H(\mathbf{s}) -  \overline{M} (\mathbf{s},\mathbf{s}^{(m)})\big) \notag \\
\hspace*{-1mm}&&\hspace*{4mm}\mbox{s.t.} \quad  \mbox{C1--C3},\widetilde{\mbox{C4}},\widetilde{\mbox{C5}},\mbox{C6b}, \mbox{C7--C21}.
\end{eqnarray}
In problem \eqref{sdr-dc}, the objective function and all constraints are convex,  such that the problem becomes a convex SDP which can be solved efficiently by standard convex program solvers such as CVX \cite{website:CVX}.
Besides, the tightness of the adopted SDP relaxation is verified in the following theorem.
\begin{Thm}\label{thm:rankone_condition}
If $P_{\mathrm{max}}^{\mathrm{DL}}>0$, the optimal beamforming matrix $\tilde{\mathbf{W}}_{k,j}^i$ in the relaxed problem in \eqref{sdr-dc} is a rank-one matrix.
\end{Thm}
\emph{\quad Proof: } The proof is omitted due to the space limitation\footnote{Theorem 1 can be proved using a similar approach as in the Appendix of \cite{sun17WSAsecureFD}.}. \hfill\qed

The optimal value of problem \eqref{sdr-dc} serves as a lower bound of \eqref{dc-penalty-pro}.
Then, we employ an iterative algorithm to tighten the obtained lower bound as summarized in \textbf{Algorithm 1}.
By solving the convex lower bound problem in \eqref{sdr-dc}, the proposed iterative scheme generates a sequence of feasible solutions $\mathbf{s}^{(m+1)}$, $\tilde{\mathbf{W}}^{(m+1)}$, $\tilde{\mathbf{Z}}^{(m+1)}$, and $\tilde{\mathbf{p}}^{(m+1)}$.
It can be shown that the proposed suboptimal iterative algorithm converges to a locally optimal solution of \eqref{dc-penalty-pro} with polynomial time computational complexity \cite{dinh2010local}.

\section{Simulation Results}

\begin{table}[t]\vspace*{-0mm}\caption{System parameters employed in simulations.}\vspace*{-1mm}\label{tab:parameters} 
\newcommand{\tabincell}[2]{\begin{tabular}{@{}#1@{}}#2\end{tabular}}
\centering
\begin{tabular}{|l|l|}\hline
\hspace*{-1mm}Carrier center frequency and bandwidth & $2$ GHz and $5$ MHz \\
\hline
\hspace*{-1mm}Number of subcarriers, ${N_{\mathrm{F}}}$ & $64$  \\
\hline
\hspace*{-1mm}Bandwidth of each subcarrier & $78$ kHz \\
\hline
\hspace*{-1mm}Path loss exponent and reference distance   &  \mbox{$3.6$} and  $15$ meters  \\
\hline
\hspace*{-1mm}BS antenna gain and SI cancellation constant, $\rho$ &  \mbox{$10$ dBi} and  \mbox{$-100$ dB}  \\
\hline
\hspace*{-1mm}Maximum tolerable data rate, $R_{\mathrm{tol}_{k}}^{i\mathrm{DL}}$ and $R_{\mathrm{tol}_{j}}^{i\mathrm{UL}}$ &  $0.3$ bits/s/Hz  \\
\hline
\hspace*{-1mm}Maximum transmit power for UL users, $P_{\mathrm{max}_j}^{\mathrm{UL}}$ &  \mbox{$18$ dBm}   \\
\hline
\hspace*{-1mm}Penalty factor $\eta$ for \textbf{Algorithm 1} &  $10 \hspace*{-0.4mm} \log_2(1\hspace*{-0.8mm}+\hspace*{-0.8mm}P_{\mathrm{max}}^{\mathrm{DL}}\hspace*{-0.4mm}/\hspace*{-0.4mm}\sigma_{\mathrm{UL}}^2\hspace*{-0.8mm})$  \\
\hline
\end{tabular}
\vspace*{-4mm}
\end{table}

In this section, we investigate the performance of the proposed resource allocation scheme through simulations. The adopted simulation parameters are given in Table \ref{tab:parameters}.
We consider a single cell where the FD BS is located at the center of the cell.
The users and the potential eavesdroppers are randomly and uniformly distributed between the reference distance and the maximum service distance of $500$ meters. The weights of all users are set as 1, i.e., $w_k=\mu_j=1, \forall k,j$. The small scale fading of the DL channels, UL channels, CCI channels, and eavesdropping channels is modeled as independent and identically Rayleigh distributed. The multipath fading coefficient of the SI channel is generated as independent and identically distributed Rician random variable with Rician factor $5$ dB. The noise powers of the DL users, the FD BS, and the potential eavesdroppers are set to $-110$ dBm. The maximum number of iterations $I_{\mathrm{max}}$ for \textbf{Algorithm 1} is set to $2N_{\mathrm{F}}$.

For comparison, we consider two baseline schemes. For baseline scheme 1, we adopt maximum ratio transmission beamforming (MRT-BF) for DL transmission where the direction of beamformer $\mathbf{w}_k^i$ is identical with the channel vector $\mathbf{h}_k^i$. Then, we jointly optimize $\mathbf{Z}^i$, $P_j^i$, and the power allocated to $\mathbf{w}_k^i$. For baseline scheme 2, we adopt an isotropic radiation pattern for $\mathbf{Z}^i$ and optimize $\mathbf{w}_k^i$ and $P_j^i$.

\begin{figure}
\centering\vspace*{-3mm}
\includegraphics[width=3.5in]{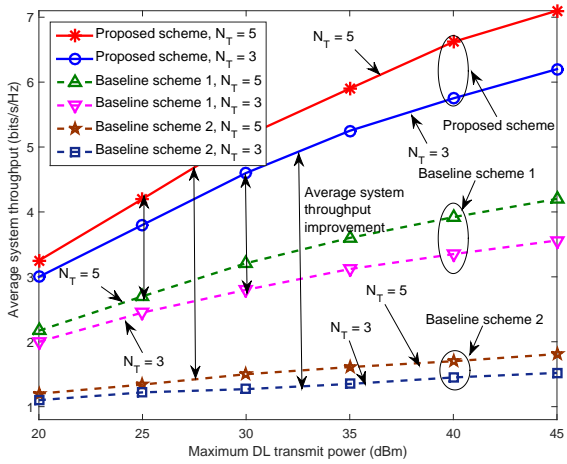}\vspace*{-3mm}
\caption{Average system throughput (bits/s/Hz) vs. the maximum DL transmit power at the FD BS (dBm), $P_{\mathrm{max}}^{\mathrm{DL}}$, for different resource allocation schemes. The double-sided arrows indicate the performance gains of the proposed optimal scheme compared to the baseline schemes.}
\label{fig:throughput_vs_power}\vspace*{-4mm}
\end{figure}

\vspace*{-0mm}
Figure \ref{fig:throughput_vs_power} illustrates the average system throughput versus (vs.) the maximum DL transmit power at the FD BS, $P_{\mathrm{max}}^{\mathrm{DL}}$, for $K=4$ DL users, $J=4$ UL users, and $M=2$ potential eavesdroppers.
As expected, the average system throughput of the proposed scheme increases monotonically with the maximum transmit power $P_{\mathrm{max}}^{\mathrm{DL}}$. Besides,
the average system throughput of the proposed scheme improves with increasing number of antennas $N_{\mathrm{T}}$ at the FD BS. This is because the extra degrees of freedom offered by additional antennas facilitate more precise and efficient information beamforming and AN generation. On the other hand, both baseline schemes achieve a significantly lower average system throughput compared to the proposed scheme.
For baseline scheme 1, since the fixed information beamforming design causes severe information leakage, more power is needed for AN generation to interfere the potential eavesdroppers, which degrades the system performance. For baseline scheme 2, the fixed AN design cannot provide reliable communication security and interferes DL transmission and UL reception severely.

\begin{figure}
\centering\vspace*{-3mm}
\includegraphics[width=3.5in]{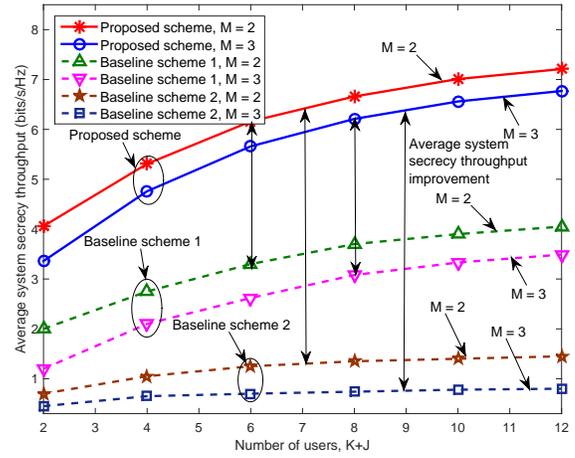}\vspace*{-3mm}
\caption{Average system secrecy throughput (bits/s/Hz) vs. the total number of users, $K+J$, for $P_{\mathrm{max}}^{\mathrm{DL}}=45$ dBm. The double-sided arrows indicate the performance gains of the proposed optimal scheme compared to the baseline schemes.}
\label{fig:secrecy_rate_vs_usernum}\vspace*{-4mm}
\end{figure}

\vspace*{-0mm}
Figure \ref{fig:secrecy_rate_vs_usernum} illustrates the average system \textit{secrecy} throughput vs. the number of users for a maximum transmit power of $P_{\mathrm{max}}^{\mathrm{DL}}=45$ dBm at the FD BS and $N_{\mathrm{T}}=5$. We assume that the numbers of DL and UL users are identical, i.e., $K=J$. As can be observed, the average system secrecy throughput for the proposed scheme and the baseline schemes increases with the number of users since these schemes can exploit multiuser diversity.
However, the average system secrecy throughput of the proposed scheme grows faster with the number of users than that of the baseline schemes. This is because the proposed scheme is able to fully exploit the spatial degrees of freedom of the considered system by optimizing both the information beamforming and the AN generation, which results in a higher multiuser diversity gain compared to the baseline schemes, which optimize either the information beamforming (baseline scheme 2) or the AN generation (baseline scheme 1) but not both.
Besides, both the proposed scheme and the baseline schemes achieve a lower average system secrecy throughput when there are more potential eavesdroppers in the system. In fact, for a larger $M$, the BS has to dedicate more radio resources to interfering the potential eavesdroppers and reducing the information leakage.

\vspace*{-0mm}
\section{Conclusion}
In this paper, we studied the resource allocation algorithm design for secure FD OFDMA systems. The maximization of the weighted system throughput was formulated as a mixed combinatorial non-convex optimization problem for joint precoding and power and subcarrier allocation algorithm design. The considered resource allocation framework limits the information leakage to guarantee secure DL and UL transmission.
A suboptimal iterative algorithm having polynomial time computational complexity was developed.
Simulation results revealed that the proposed suboptimal resource allocation scheme achieves a significantly higher performance than two baseline schemes.

\vspace*{-0mm}
\bibliographystyle{IEEEtran}
\bibliography{secure_FD_OFDM}

\end{document}